\documentclass[sigconf]{acmart}

\AtBeginDocument{%
  \providecommand\BibTeX{{%
    \normalfont B\kern-0.5em{\scshape i\kern-0.25em b}\kern-0.8em\TeX}}}

\setcopyright{acmcopyright}
\copyrightyear{2023}
\acmYear{2023}
\acmDOI{10.1145/3616195.3616209}

\acmConference[AM '23]{Audio Mostly}{August 30 -- September 01,
  2023}{Edinburgh, UK}
\acmPrice{15.00}
\acmISBN{979-8-4007-0818-3}

\begin{document}

\title[Stringesthesia: Dynamically Shifting Musical Agency Between Audience and
Performer Based on Trust]{Stringesthesia: Dynamically Shifting Musical Agency Between Audience and
Performer Based on Trust in an Interactive and Improvised Performance}


\author{Torin Hopkins*‡, Emily Doherty‡†, Netta Ofer*, Suibi Che Chuan Weng*, Peter Gyory*, Chad Tobin*, Leanne Hirshfield‡†, Ellen Yi-Luen Do*‡†}
    \affiliation{
    \department{ATLAS Institute*, Institute of Cognitive Science‡, Computer Science†}
      \institution{University of Colorado Boulder}
      \country{Boulder, CO, USA}
    }
    \email{Torin.Hopkins@colorado.edu}

\renewcommand{\shortauthors}{Hopkins, et al.}

\begin{abstract}
 This paper introduces Stringesthesia, an interactive and improvised performance paradigm. Stringesthesia uses real-time neuroimaging to connect performers and audiences, enabling direct access to the performer's mental state and determining audience participation during the performance. Functional near-infrared spectroscopy (fNIRS), a noninvasive neuroimaging tool, was used to assess metabolic activity of brain areas collectively associated with a metric we call “trust”. A visualization representing the real-time measurement of the performer’s level of trust was projected behind the performer and used to dynamically restrict or promote audience participation: e.g., as the performer’s trust in the audience grew, more participatory stations for playing drums and selecting the performer’s chords were activated. Throughout the paper we discuss prior work that heavily influenced our design, conceptual and methodological issues with using fNIRS technology, and our system architecture. We then describe feedback from the audience and performer in a performance setting with a solo guitar player.
\end{abstract}







\maketitle

\section{Introduction}
Trust is a crucial component of promoting performer-audience engagement, as both parties must express some level of confidence to carry out a successful performance \cite{mccormack_hutchings_gifford_yee-king_llano_dinverno_2020}. While the interplay of trust has been explored in some interactive musical performances \cite{mccormack2019silent, mccormack_hutchings_gifford_yee-king_llano_dinverno_2020}, much of the relevant trust literature comes from human-automation interaction studies \cite{eloy2022using,bobko2022human}. These studies evaluate trust often through the manipulation of an artificial agent's characteristics such as transparency and reliability. In particular, when agents provide unreliable or incomplete information, their human collaborators are likely to ignore the information and rely solely on their own experience, resulting in a lack of trust in the agent. Conversely, when agents provide information that is overly reliable, humans may become complacent and over-rely on the agent, resulting in a lack of vigilance and degraded performance \cite{hoff2015trust, glikson2020human, eloy2022using}. Communication or lack thereof between team members can reveal the level of trust, which can greatly impact the effectiveness of collaboration.

Similarly, in audience-interactive musical improvisation, trust is built through communication between musicians and audience members, as well as  openness to the creative process \cite{jeaton2014, leslie2014measuring}. However, establishing trust can be a delicate balance. Unreliable (e.g. non-interpretable rhythms) or overly reliable (e.g. static and unchanging rhythms) musical performance can  negatively impact the outcome of the musical collaboration \cite{mccormack2019silent,mccormack_hutchings_gifford_yee-king_llano_dinverno_2020}.

To explore trust and interactivity in a musical performance setting, some musicians have included the audience in interactive and improvised performance paradigms, where the audience contributes music notation or influences the direction of the music with semiotic information \cite{Lee2013, slee12014, NIME22_3, NIME20_11, Fan2013a, Lee2013c, Hindle2013, Weitzner2012}. For example, an audience could control the speed or volume of an improvising musician's music based on physical controllers positioned in the audience. However, there is a trade-off between how much musical agency, defined as control over the musical content, to give the audience vs. the performing musician/s. Determining musical agency between audience and performer significantly influences the outcome of the musical performance in interactive and collaborative performance paradigms \cite{mudd2019material, brown2019case, hein2019groove}.

Bio-sensing techniques have been used to enhance collaboration through transparency in both human-AI interaction studies and interactive musical performance. For example, bio-sensing has been used to display an autonomous driving system's level of certainty, which can inform the driver to take over control before a collision event \cite{kunze2019automation}. In musical performance, bio-sensing has been used to approximate trust and inform performers of an AI-musician's confidence in their musical output \cite{mccormack2019silent}. Additionally, bio-sensing techniques have been used to involve the audience in the performance, creating a unique sense of intimacy and engagement between the performer and the audience \cite{Fan2013a, mullen2015mindmusic, jeaton2014, leslie2014measuring}.

A commonly used technique to analyze neural signals in performances is electroencephalography (EEG) \cite{lucier2012chambers,Fan2013a,leslie2014measuring, mullen2015mindmusic}. However, fNIRS  has greater spatial resolution than EEG, offering more information about the localization of neural activity associated with cognitive states like trust \cite{perello2022using}. The benefits of fNIRS have also been underexplored in creative musical settings and further exploration will uncover more about its benefits and drawbacks \cite{Yuksel2019}.

\begin{figure*}
  \includegraphics[width=\textwidth,height=5.5cm]{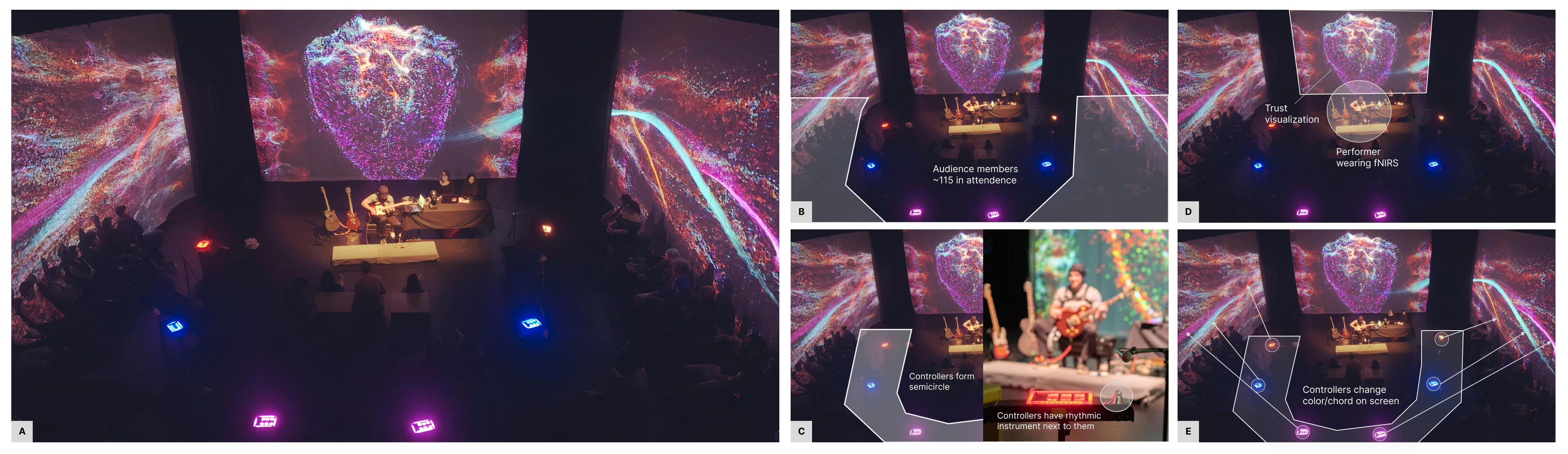}
  \caption{A) Shows the full performance area in Stringesthesia. B) Location of the 115 audience members during performance. C) Shows audience controllers and associated instruments at each controller. D) Location of performer and visualization of trust. E) How the color/chord controllers affected colors of immersive content and trust visualization.}
  \label{fig:sys_overview}
\end{figure*}

In this paper, we present Stringesthesia, a collaborative musical improvisation that explores intimacy between the audience and performer that is not as prevalent in traditional performance paradigms (see Figure \ref{fig:sys_overview}). Stringesthesia measures the performer's trust level in real-time using fNIRS, which feeds back to affect the aspects of the performance the audience can control, such as selecting chords the musician has to play and how many audience members can play drums. This feedback is designed to dynamically tune the level of agency the audience or performer have during the performance, such that when trust is high audience members have more agency, and when trust is low the performer has more agency. We present the development, implementation, design process, and performance of Stringesthesia, as well as reflect on feedback from the audience and performer for improvements in future performances.

\section{Related Works}
The following sections outline prior work that heavily influenced the components of Stringesthesia.

\subsection{Neural Signals in Interactive Musical Performance}
Biologically-inspired performance paradigms include using electroencephalogram (EEG) and pulse oximetry to measure neural and cardiac signals that were used to guide the tempo and bidirectional communication between audience and performer \cite{Fan2013a}. For example, the installation \textit{Ringing Minds}, explores concepts in multi-musician performance in active and imaginative listening. This promotes the creation of a unique performance whereby collective brain responses of four audience members interact to inform a spontaneously generated musical piece \cite{mullen2015mindmusic}.

Another example of musician-audience interaction using EEG is the installation \textit{Spukhafte Fernwirkung} (Spooky Action at a Distance) \cite{mullen2015mindmusic}. This installation focuses on providing musical ideas for an improvising piano player by reading EEG signals from an audience member. These examples and more inform the way that players, audience members, and installation participants can co-create music using the information provided by the brain of either the performing artist/s or audience members.

An example of a live collaborative improvisation using neurofeedback is Eaton’s performance called, \textit{The Space Between Us} \cite{jeaton2014}. In this performance EEG signals are read from an audience member and a performer. Emotional correlates in the EEG signal are converted into musical phrases played by a computer and a musical score that is presented to a vocalist and piano performer. During the performance, both the performing vocalist's and the audience member’s EEG signals inform the selection of musical passages and seek to illuminate the similarities and differences in their respective emotional states during a live performance.

\subsection{Musical Agency}
Musical agency is a critical concept in interactive music systems, referring to an individual's ability to control and influence musical outcomes through their actions, decisions, and intentions. It involves a sense of empowerment and responsibility over the music-making process, as well as the ability to shape the musical expression and meaning. In the context of interactive dance systems, for example, dancers with high musical agency are able to create and manipulate musical sounds through their movements and gestures, and can understand how their actions contribute to the overall musical outcome \cite{brown2019case}.

Musical agency has been explored in contexts such as digital musical instrument design, collaborative music making, and interactive performance. We focus on musical agency given to audiences in performance in its most commonly used form, music notation.

\subsubsection{Music Notation in Interactive Performance for Giving Audience Members Agency}
Audience-performer interactivity has a rich history beginning largely with audience contributed music notations, as well as mobile and web applications \cite{Lee2013, slee12014, NIME22_3, NIME20_11, Fan2013a, Lee2013c, Hindle2013, Weitzner2012}. These systems use audience-contributed classical music notation, interfaces for audio-based snippets, or translate more accessible media such as text into classical or other forms of musical notation for the performer \cite{Baird2005, Lee2012a, Dahl2011, jeaton2014}. Musical notation is a powerful tool for the audience to provide information to the performing musician because of its direct relevance to the content of the performed music. Whether the notation is informed by motion \cite{Bouchard_tenor2019} or another technology, music notation can inspire new musical directions or directly translate into performed music.

Inspired by these examples, Stringesthesia utilizes color as a mechanism of delivering musical information to the performer. The performer has a trained lexical understanding of the color-chord relationship and therefore utilizes the information provided in the audience designed visualization provided for the performing artist.

\subsection{Trust and Affective State Modeling}
Trust is a complex, multidimensional construct that has been studied extensively in contexts such as interpersonal interactions, relationships, and in interactions with automation \cite{fett2014default}. Recent research on trust in robotic, virtual, and embedded artificial intelligent systems highlights the importance of reliability in building trust \cite{glikson2020human,hoff2015trust}. Reliability is the extent to which a behavior is accurate and consistent, which can be seen as a situational factor rather than a design or experimental factor. Changes in reliability are perceived as an inherent consequence of imperfect AI \cite{glikson2020human}. Trust has been described as one’s willingness to rely on an external source, and often demonstrates positive expectations of the trusting party \cite{rousseau1998not}.

In the context of musical performance, reliability can be seen as a key component of trust-building, as unpredictable or unreliable musical performance can negatively impact the outcome of musical collaborations. Therefore, musicians must strive to establish a high level of reliability in their musical performance to build trust with their fellow musicians and audience members \cite{mccormack2019silent,mccormack_hutchings_gifford_yee-king_llano_dinverno_2020}. This can be achieved through consistent and accurate execution of musical elements such as rhythm, melody, and harmony, as well as through transparent communication and openness to feedback from collaborators and audiences.


If the performing audience members are too unreliable, for example, if their rhythms are non-interpretable or their musical contributions are disjointed and disconnected, the performer may become disengaged or lose trust in the performance. On the other hand, if the performing audience members are too rigid and unchanging, the performer may become bored or feel that the performance lacks creativity and spontaneity. Therefore, trust in musical performance requires a balance between reliability and unpredictability, as well as effective communication and openness to the creative process between performers and audience members.



\subsubsection{fNIRS Data}
Functional Near-Infrared Spectroscopy (fNIRS) is a non-invasive and lightweight neuroimaging modality that measures changes in hemoglobin concentration within the brain using infrared light \cite{ayaz2022optical}. It offers several advantages over similar modalities such as EEG and functional magnetic resonance imaging (fMRI). It is non-invasive which enables examination in real-life, naturalistic settings. Compared to other forms of neuroimaging, it is resistant to motion and light artifacts. fNIRS is also more safe and cost-effective compared to fMRI, and has greater spatial resolution than EEG \cite{hirshfield2019toward}.
Following neuronal activation within a region of the brain, oxygenated hemoglobin travels to that brain area. Infrared light is used to detect changes in hemoglobin concentration (oxygenated and deoxygenated), which thus can be used to infer brain activity \cite{liu2015inferring}. fNIRS has been shown to be suitable for measurement during complex cognitive and physical tasks such as yoga \cite{dybvik2021real}, dance \cite{noah2015fmri}, and classical music performance \cite{vanzella2019fnirs}, therefore we used fNIRS during the Stringethesia performance for live neural recording.


\subsubsection{Brain Regions Associated with Trust}



\begin{figure}[htbp]
	\centering
		\includegraphics[width=1\columnwidth]{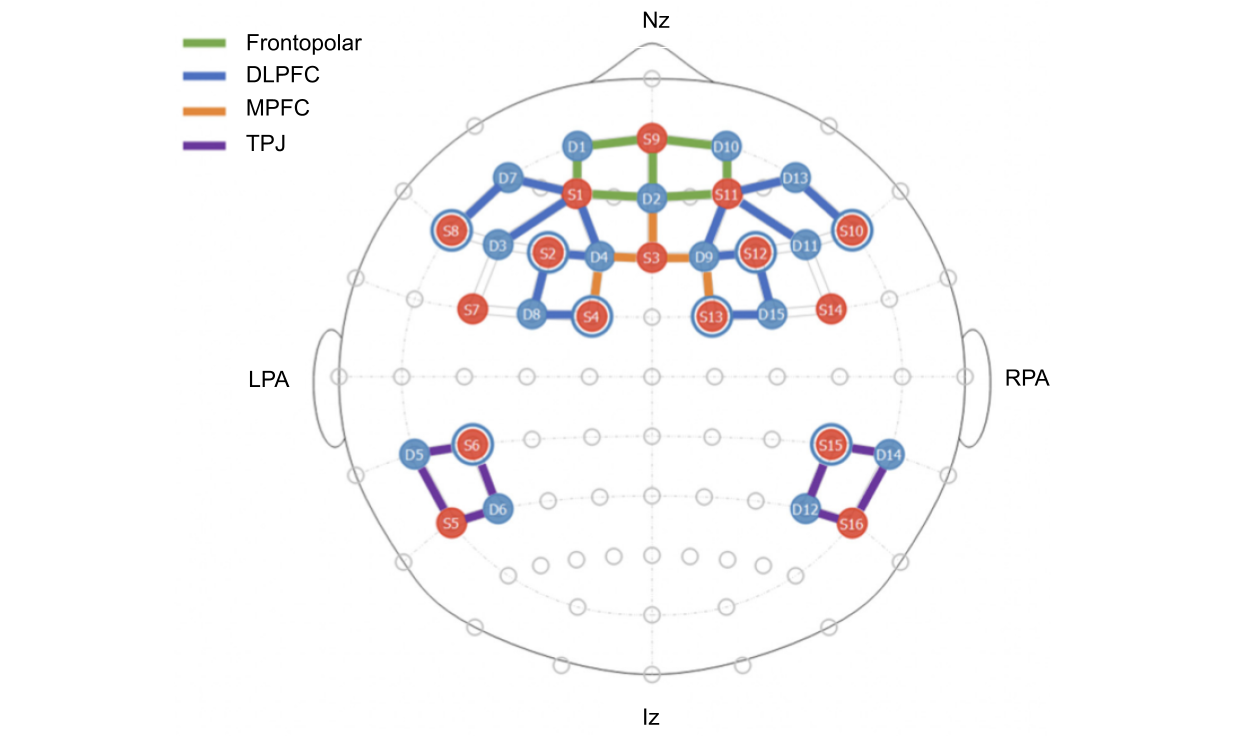}
	\caption{Top-down view of the sensor and detector arrangement for the measurement of trust (S = sensor, D = detector). LPA and RPA show left and right side of the head respectively. Iz and Nz represent the back and front of the head. Colored lines discriminate brain regions.}
	\label{fig:HAT montage}
\end{figure}

As fNIRS is a newer neuroimaging technique than comparable technologies like fMRI and EEG, the literature using fNIRS to measure correlates of trust is limited~\cite{hirshfield2019toward, eloy2022using, bobko2022human}. However, since fNIRS and fMRI both measure the hemodynamics of the brain and have been proven to be correlated \cite{liu2015inferring}, we can draw from the extensive literature studying trust using fMRI (as summarized by~\cite{hirshfield2019toward}). The main regions of interest (ROIs) implicated in decision-making and trust include the frontopolar area (FPA), medial prefrontal cortex (MPFC), dorsolateral prefrontal cortex (DLPFC), and the bilateral temporoparietal junction (TPJ)~\cite{hirshfield2019toward, eloy2022using} (see Figure \ref{fig:HAT montage}). The regions have often been implicated in Theory of Mind reasoning, a research paradigm evaluating how one attributes thoughts, intentions and beliefs to others~\cite{sebastian2012neural}. These areas are often activated during social interactions, evaluation of self and others, decision-making, and determining when to trust~\cite{sebastian2012neural, mahy2014and, eloy2022using}.

Trust was further quantified using pilot data collected prior to the performance in which the performer played both cooperatively and competitively with another musician. We found that tuning the signal present in brain regions of the trust network yielded more reliable readings of trust. The pilot data was used to inform the weighting of the respective brain regions (TPJ, FPA, MPFC, DLPFC). TPJ was given a weight of 2, meaning that these bilateral regions had greater weight in our quantification of trust. FPA was given a weight of 0.5 because FPA is heavily associated with mental workload and has been implicated in less transparent situations \cite{eloy2022using}. The other ROIs, DLPFC and MPFC, were given a weight of 1 as we hypothesize they plan a similar role in trust.

Stringesthesia draws from these paradigms by using the fNIRS neuroimaging technique to further connect the audience and performer. fNIRS is well-suited as compared to other technologies such as EEG to accurately measure affective states such as trust and complex task environments, but to the best of our knowledge there are no examples of fNIRS used for trust measurement in musical performance. Our use of fNIRS extends this concept by conveying the internal state of performers' trust to the audience and tuning the musical agency given to the audience based on the performer's sense of trust--further connecting the actions of the audience to the outcome of the performance. Because the performer has no rehearsed material the sense of trust between the performer and audience is essential for a positive musical outcome.

\section{Methods}
The performance took place on the campus of the University of Colorado, Boulder's black box theater. Equipped with six projectors, stage lights, scrum curtains, and a large rectangular floor space (approx. 2700 square feet). The concert lasted for 2 hours and 15 minutes divided by a 15 minute intermission with a total of 115 audience members in attendance.

\begin{figure}[htbp]
	\centering
		\includegraphics[width=1\columnwidth]{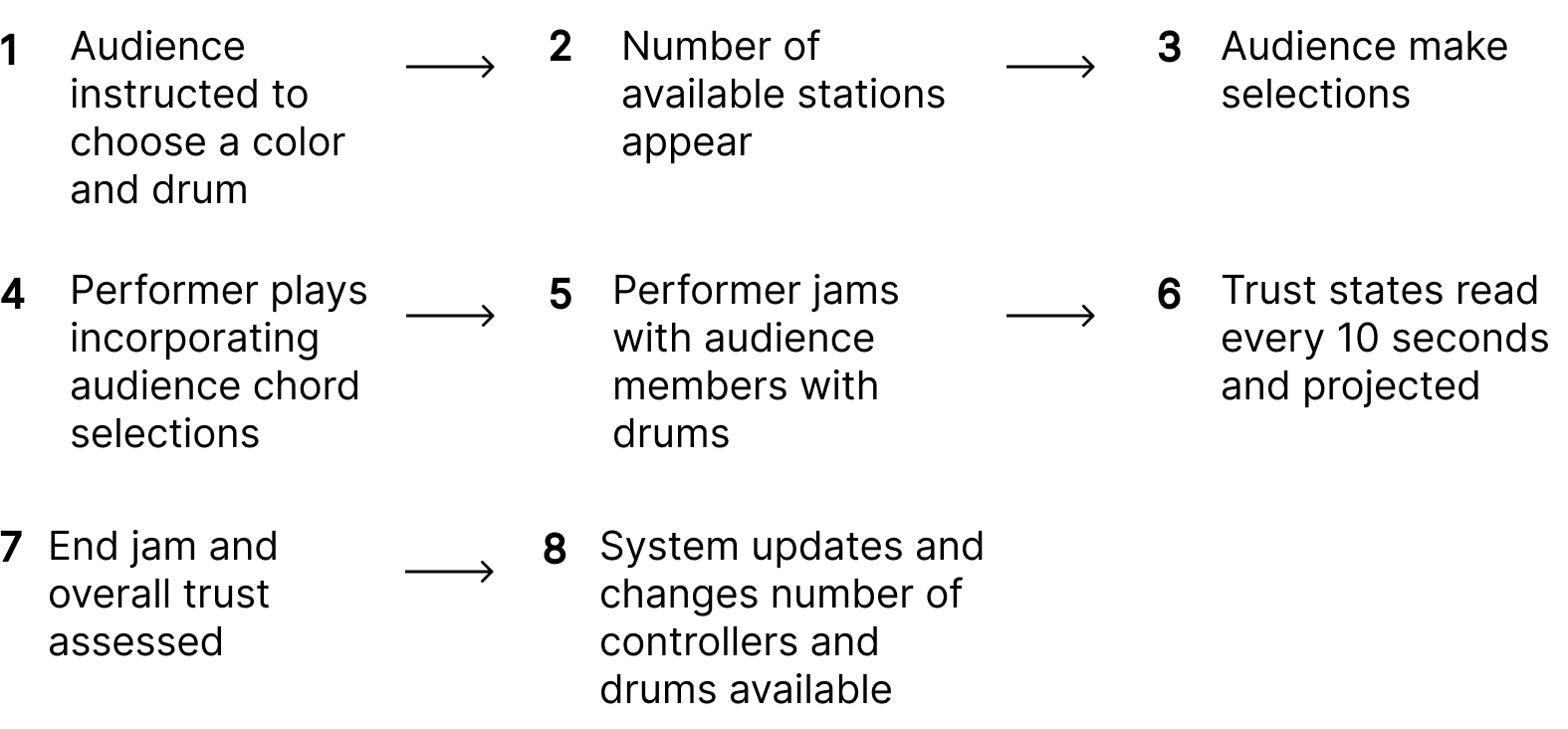}
	\caption{Demonstrates the sequence of events during performance. Cycle is repeated until performance is over. As jam sessions become longer, blood-oxygen levels can saturate. Thus, steps 7 and 8 must account for blood-oxygen levels to normalize.}
	\label{fig:ET}
\end{figure}

The timeline of events was constructed so that audience members had a chance to move about the space, switch instruments, choose colors, and interact in different ways throughout the performance. Limitations of the fNIRS readings required us to divide the performance into sets, allowing the oxygenated hemoglobin signal from the fNIRS device to return to baseline and enabling a more effective reading of trust (see Figure \ref{fig:ET}).


\subsection{Measuring and Reporting Trust Levels}
Blood-oxygen levels measured by the fNIRS device (NIRSport 2) were live-streamed via Lab Streaming Layer to MATLAB. The device had 50 channels spanning across the four aforementioned ROIs (TPJ, FPA, DLPFC, and MPFC). Lab Streaming Layer parsed the data stream and calculated the relative oxygenated blood density using the modified Beer-Lambert law equation \cite{jacques2013optical}. It was then sent to MATLAB on a local device which ran further processing on the incoming data stream. For the performance, we employed a sliding window average with a window size of 200 data points and subtracted this figure from a baseline calculated at the beginning of each jam session. The average oxygenated blood density levels were calculated at a rate of about 0.5-2Hz. Jam sessions were variable in length (2-9 minutes) with a 1-3 minute resting period in between to ensure return of signal to baseline.

\subsection{Color/Chord Selection Controllers for Musical Agency of the Audience}
Chord selection controllers were provided to the audience to give the audience more agency over the performance. When trust levels were high, more controllers to select chords that the performer had to use in the jam became active for the audience.

\begin{figure}[htbp]
	\centering
		\includegraphics[width=1\columnwidth]{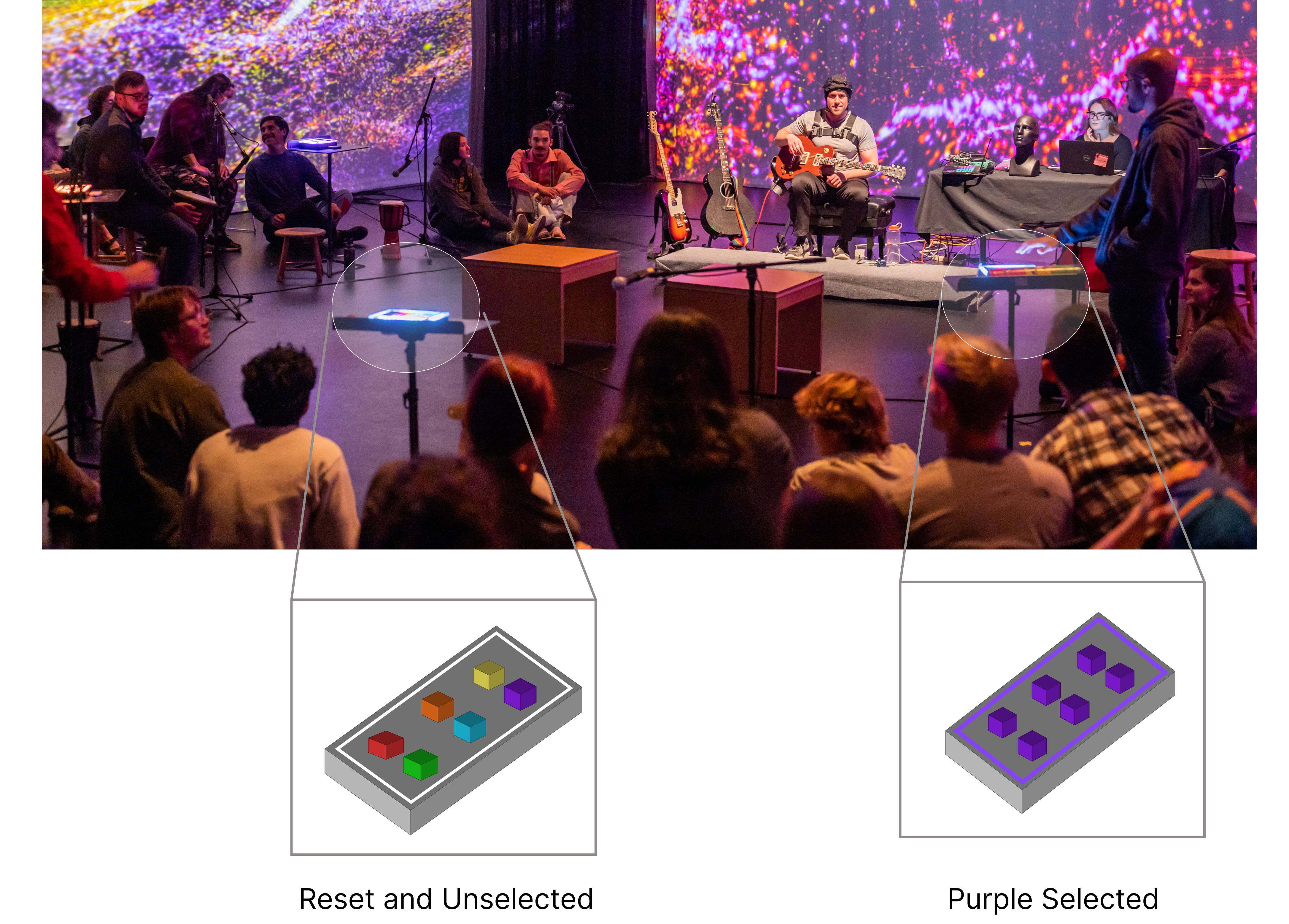}
	\caption{Picture demonstrates a controller that has been reset and one that has been assigned a color. When reset, six buttons show 6 different colors from which the audience can select a color. All colors correspond to chords the performer has memorized.}
	\label{fig:controller2}
\end{figure}

In order to bridge the gap in musical knowledge between the audience and the performer, color was used to communicate musical information. We placed 6 controllers in the performance space. Each had six different colors that the audience members could select (see Figure \ref{fig:controller2}, \ref{fig:sys_overview}, and \ref{fig:crowd}). The performance consisted of 14 different “sets” ranging from 2-9 minutes and had 1-3 minutes in between. Before each set the audience was invited to select a color. Audience members were free to select any color they wished. For the audience member, this may have been an aesthetic choice, an exploration of how the color affects the music, or any other reason the audience member may have had. Selecting a color caused the controller to change its LEDs to match the selection, and sent the selected color to the visualization system which reflected their choice on-screen.


Six colors were interpreted as chords by the performer who was trained to establish a connection between the colors and chords present in a major key. The colors were presented in a spectral order (red, orange, yellow, green, blue, purple) and represented major scale chords (I, ii, iii, IV, V, vi) respectively to inform the extemporaneous improvisation (see Figure \ref{fig:controller2}). Color choices were inspired by common mappings used in color notations between colors and notes in a major scale \cite{jamstation, Hopkins_tenor2020,hooktheory}.


\subsection{Interaction Design}
\begin{figure}[htbp]
	\centering
		\includegraphics[width=1\columnwidth]{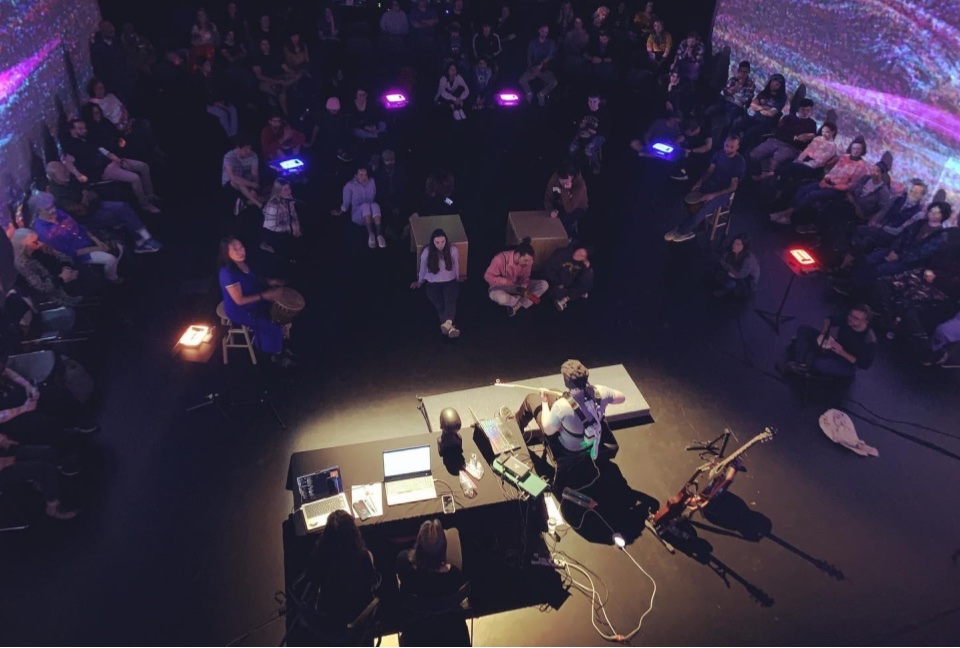}
	\caption{Shows a top-down view of the performance.}
	\label{fig:crowd}
\end{figure}

We made experience design decisions to encourage audience participation by ensuring easy access to color controllers and instruments. Four seating areas, including one in the center of the room, aimed to foster closeness between the audience and performer. The instruments were rhythmic for easy accessibility, with limited seating nearby to promote movement and sharing (see Figure \ref{fig:sys_overview} and \ref{fig:crowd}).

The performer, guitars, and fNIRS were positioned at the front, with researchers and computers visible to the audience. A short introduction explained the performance structure, and color controllers reset in between sets to invite the audience to choose again.

\subsection{Aesthetic Decisions}
Throughout the performance, we made aesthetic decisions to highlight metaphors within the system's design, focusing on visualization components and chord selection controllers for the audience.

Visualizations consisted of connection lines, representing diffusion tensor imaging, and particles resembling cell nucleus fluorescent staining techniques. We chose an orb animation displaying complex hive or flocking dynamics to signify the connection between the audience and performer, emphasizing the collaborative and improvisational nature of the performance.

Chord selection controllers, placed on music stands, symbolized the audience's role in notating music for the performer, even though the music wasn't traditional. The controllers resembled modern MIDI button pads to provide a familiar yet non-traditional feeling, indicating the performance's unconventional nature.

\section{System Architecture}
In this section we describe the various components of the system and their use in the performance.

\begin{figure}[htbp]
	\centering
		\includegraphics[width=1\columnwidth]{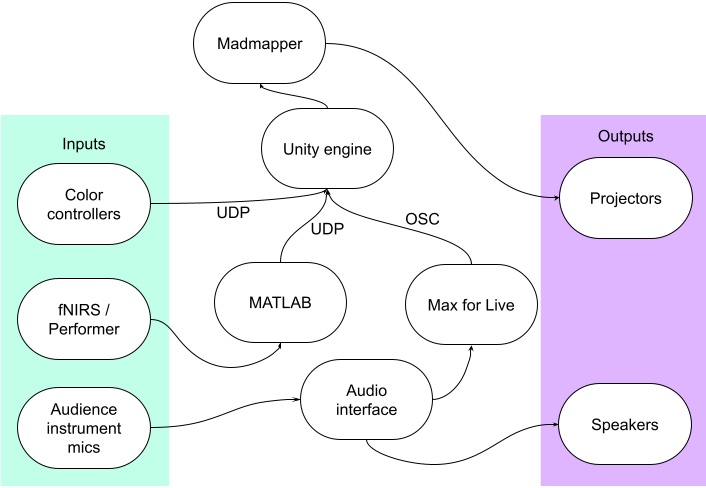}
	\caption{System connectivity diagram demonstrating inputs and outputs}
	\label{fig:SystemDiagram}
\end{figure}

\subsection{Live-Stream Pipeline}
The Stringesthesia pipeline is configured to live-stream data between the musician, audience, and a central processor that visualizes the information using projection mapping (see Figure \ref{fig:SystemDiagram}). All components are connected via WiFi to a local network transmitting messages using the Universal Datagram Protocol (UDP).

\subsection{Audio System}
Six microphones were placed near the six rhythmic instruments to create reactive visualizations for the audience. The microphones and PA amplified the sound of the instrument, as well as provided a signal to a MAX for Live 8 patch running in Ableton Live 11. The signal was passed to the backline for amplification, as well as converted to OSC messages in MAX. The volume of the audio from the receiving microphone was translated into OSC messages which encoded qualities of a visualization located near the respective drum. The audio system was used to both control the volume of the drums and provide interactive visual feedback for the audience.

\subsection{Unity Engine-Controlled Projection}
We used five projectors to create the visualizations during the performance. To connect the projectors, we used the hardware video wall controller DataPath (Fx4). After integration of the visual signals into the computer, MadMapper software (v5.1.5) was used to control the precise location of the projection.

The real-time audio-visual system was made using Unity (2021.1.7f1). The visual was made with the unity-driven plugin Visual Effect Graph (VFX Graph). The resulting visuals reacted to the fNIRS and controller signals; both sent via UDP.

\subsubsection{Visualization of Trust and Musical Action}
The level of trust corresponded to the diameter of the orb behind the performer (when the orb was small trust was low, and when the orb was large trust was high) (see Figure \ref{fig:sys_overview}). Particles appeared to be flowing to and from the audience members via a stream of diffuse particles and colored lines, connecting the performer to the audience member. The diffuse stream of colored particles converged on the orb creating a multicolored representation of trust. We chose an animation for the orb of particles that appeared to be a complex motion resembling a hive or flocking dynamics. This choice was meant to represent the connection between the audience and the performer as having a life of its own–further drawing the audience toward the shared experience of collaboration and improvisation.





\section{Feedback from Audience and Performer}
In this section we report on feedback about the experience from both the audience and performer about trust, musical agency, unexpected occurrences, and overall comments on performance.

\subsection{Audience Feedback}
Audience feedback was collected through anonymous written statements about the performance immediately after the show as well as by email followup. In our proposed paradigm, audience participation is essential to assess a sense of the performer's trust and to create a meaningful performance. Given this was not a traditional or controlled research setting, we set up a comment box outside of the auditorium along with pens and paper, prompting audience members who were interested in telling us about their experience to leave an anonymous comment.

We specifically were interested in ways to improve the performance, thoughts on perceived musical agency, and whether the audience members felt a sense of connection between each other and the performer. Some audience members reported feeling connected to each other. Three comments in particular illustrated this: 1) \textit{“This was a really cool and immersive, multisensory experience. I felt like I was really able to be a part of it and felt involved and connected to the other audience members”}, 2) \textit{“I’m feeling so inspired to jam with any and all objects and with everyone I know. Thank you!”}, and 3) \textit{“I was discussing the performance with people around me, sharing our understandings of the visualization. I was also encouraging my friend to participate by bringing her up to one of the button stations, and when she said she wanted to play an instrument but she wasn't sure if any were available. In other performances, I don't usually have as much of these discussions and interactions.”}

Some audience members noted that they did not participate musically at all during the show, stating that they would much rather watch others play music. This was observed as a continuum between those who preferred to interact and those who did not. Those that did interact described excitement to play and choose colors rather than an explicit sense of musical agency. We interpret this to mean that the ability to playfully interact without being entirely in the spotlight was exciting and interesting for many. The number of open stations was a factor that limited the number of people that could participate, but did not diminish the experience of any individual with respect to musical agency.

Some members felt the visualization of trust to be too abstract to understand what was being read. One listener stated: \textit{“Because the visualization was abstract, I wasn't very focused on interpreting the performer's trust state. The main thing I noticed was that the visualization behind the performer was usually small in the beginning of each piece, and would grow in volume over the course of the performance.}.” Another listener suggested that the \textit{“visualizations were striking”}, but wanted to know more about its theoretical underpinnings.

Others felt the visualizations were an exciting component of their experience: \textit{“I felt highly connected [to the performer] because of the immersive visuals and audio, and the visualization of the trust level gave me more of a glimpse into the performer's state.”} Another audience member said: \textit{“I was fixated on the graphics, like the orb."}

Overall many audience members indicated that their experience was in line with our interest in making a performance with over 100 people in the audience feel very intimate. One audience member reported: \textit{“The performance felt like a combination of the production quality of larger concerts I've been to (particularly due to the visual effects), with the intimacy of small performances, due to the audience participation. It was similar in the sense that despite everything going on, for me, the focus was still on the incredible music; the technology and visualization of trust levels just helped to enhance that experience. I think overall, these elements made the experience impactful.”}

\subsection{Performer Feedback}

The performer's perspective offers invaluable insights into the dynamics of the interactive and improvised performance. The performer noted the impact of the audience's emotions on the musical experience, stating, "\textit{The audience's excitement and nervousness to participate added an extra dimension to the performance}". This highlights the importance of considering audience emotions when designing interactive performances, as these emotions can directly impact the overall atmosphere \cite{van2011performers, fazekas2014novel}.

In instances where there were fewer controllers (low trust levels), the musician reported, "\textit{I had more musical freedom}". This observation suggests a trade-off between audience interaction and artistic control. Despite some audience members playing rhythmic instruments even when their station was deactivated, the performer appreciated the spontaneous nature of these interactions, saying, "\textit{It added an interesting element to the performance}".

The performer also noted that at times, the audience reclaimed agency by stomping feet or prolonging beats beyond the musician's intent. Reflecting on these actions, the performer said, "\textit{These interactions fostered a high level of engagement between the audience and me, making the performance more immersive}".

Overall, the performer described the concert as "\textit{a fun, engaging, and immersive experience}". They acknowledged that there was sufficient space between sets for brain levels to desaturate, allowing for dramatic musical changes, and remarked, "\textit{This balance contributed to the success of the interactive and improvised performance}".

The performer's feedback provides a comprehensive understanding of the various aspects of the interactive and improvised musical performance. It highlights the importance of balancing audience interaction with artistic control, as well as the need to consider audience emotions when designing interactive performances. By considering the performer's perspective, future academic musical performance papers can better evaluate and enhance the concert experience for both the audience and the musicians involved.
\section{Discussion}
The interactive and improvised performance of Stringesthesia explored the intricate balance of trust between the performer and audience. The audience's enthusiasm and apprehension when participating added a unique dynamic to the overall experience, emphasizing the importance of considering audience emotions in the design of interactive performances. The audience's involvement in the performance, along with the use of fNIRS to measure the performer's trust levels in real-time, provided a rich context for understanding interactions between the performer and audience.

The feedback from both the audience and the performer reveals some valuable insights. The audience felt a sense of connection with each other and the performer, facilitated by the immersive visuals and interactive nature of the performance. However, some participants found the trust visualization abstract and difficult to interpret, suggesting a need for a more comprehensible representation or additional explanation before the performance. This could be addressed by providing a more in-depth program that outlines the role of trust and its visualization in the performance.

The performer's feedback confirms the delicate balance between audience interaction and artistic control. They appreciated the musical freedom they experienced during moments of low trust levels, as well as the spontaneous nature of abundant audience involvement. The audience's active participation in the performance, such as stomping feet and prolonging beats, fostered a high level of engagement between the performer and audience members.

The diverse preferences and experiences of the audience highlight the importance of accommodating different levels of engagement. Some audience members preferred not to participate musically, enjoying watching others play instead, while others found excitement in playfully interacting without being completely in the spotlight. The limited number of open stations constrained the number of participants but didn't diminish any individual's experience concerning musical agency.

Stringesthesia presents a novel performance paradigm that promotes audience participation and fosters a sense of intimacy and connection between the performer and the audience. The insights gained from the audience and performer feedback highlight areas for improvement, such as refining the trust visualization and providing a better explanation of its significance in the performance. By addressing these issues and taking into account the lessons learned from this performance, future interactive and improvised musical performances can strike the right balance between audience engagement and artistic control, resulting in a more meaningful and fulfilling experience for all involved.

\section{Limitations and Future Work}
When interpreting fNIRS data, it is important to note that hemodynamic signal lags behind neuronal activation. The lag (approx. 3-5 sec \cite{west2019bold}) imparted by the technology influenced the performance structure. Both real-time measurements of trust (using a sliding window average) and an overall average of trust at the end of every set were assessed to account for both hemodynamic signal saturation and the signal delay. EEG has better temporal resolution in this sense, but less spatial specificity offered by fNIRS. A future study could take advantage of the complimentary features of these modalities and use concurrent fNIRS-EEG recording during a performance to measure performer-audience trust.

Additionally, reading neural data from the crowd would add a level of complexity that may inspire new musical action. The cost of fNIRS machines hinders this desire, but there are other biological sensors that can (though less accurately) approximate trust if we explore this option.

A musical limitation was imposed by only having 6 colors used to represent chords in a Western music derived major key. In future performances this can be extended to use a larger range of chords, chord qualities, musical style, as well as color palettes. Additional investigations as to how much influence the audience has over the performance is also of interest to us in the future.

\section{Conclusion}

Stringesthesia offers an innovative performance paradigm that encourages audience participation and cultivates a sense of connection between the performer and audience members. The use of fNIRS to measure the performer's trust levels in real-time provided valuable insights into the delicate balance between audience interaction and artistic control. Feedback from both the performer and audience highlighted the importance of considering audience emotions and fostering engagement while maintaining artistic freedom. The need for clearer communication regarding the trust visualization was identified, which can be addressed in future performances. By incorporating these lessons, future interactive and improvised musical performances can create a more enriching and immersive experience for everyone involved.
\begin{acks}
{We would like to acknowledge the ATLAS Institute's B2 Center for Media, Arts \& Performance at the University of Colorado Boulder for facilitating the performance and assisting in the production of the show. We thank Sean Winters and Chris Petillo for support with audio, as well as Rishi Vanukuru for aiding with the performance. We would also like to thank the audience that came to experience the performance with us and the reviewers for their insights.}
\end{acks}

\bibliographystyle{ACM-Reference-Format}
\bibliography{am23-14}


\end{document}